# SECURE ARCADE: A GAMIFIED DEFENSE AGAINST CYBER ATTACKS

Sean Loesch, Ryan Hrastich, Jordan Herbert, Ben Drangstveit, Jacob Weber, Mounika Vanamala

Department of Computer Science, University of Wisconsin-Eau Claire, Eau Claire, Wisconsin, USA

## ABSTRACT

*In modernity, we continually receive increasingly intricate technologies that allow us to increase our lives' convenience and efficiency. Our technology, particularly technology available over the internet, is advancing at unprecedented speed. However, this speed of advancement allows those behind malicious attacks to have an increasingly easier time taking advantage of those who know little about computer security. Unfortunately, education in the computer security field is generally limited only to tertiary education. This research addresses this problem through a gamified web-based application that drives users to reach learning goals to help them become more vigilant internet users: 1. Learn and memorize general computer security terminology, 2. Become familiar with basic cryptography concepts, 3. Learn to recognize potential phishing scams via email quickly, and 4. Learn common attacks on servers and how to deal with them.*

## 1. INTRODUCTION

In the age of increasing interconnection via the internet, computer security is crucial for protecting personal data and devices from malicious threats. Modern forms of assault on our privacy, such as deepfakes, new malware, and spoofing, among others, constantly loom over our heads as we browse the internet [1], [2], [3], [4]. With over two billion personal computers and a smartphone penetration rate of 67% worldwide [5], computer security has become paramount in safeguarding data. Computer security has driven all our technological advancements [6], allowing us to increase the degree of interconnectivity and still thrive in today's digital age. However, as more individuals become connected and use technology, without a corresponding increase in knowledge about how to protect themselves, the chances of being caught off guard against cyber-attacks increase [7], [8]. Nobody who browses websites is totally safe [9], [10].

Moreover, computer security education is often limited to tertiary education [11], which not everyone can afford. This predicament leaves individuals vulnerable to online threats, especially with the rise of remote work and distance learning [12]. As a result, we have made it our mission to develop an online application that utilizes an engaging and interactive minigame system to teach computer security concepts. By increasing knowledgeability in this field, we aim to promote higher vigilance online among internet users to help them protect themselves and their digital assets. In this research, we developed a gamified, web-based application to administer this knowledge to achieve this desired effect.

We identified four key learning objectives to encapsulate computer security succinctly:

1. Learn and memorize general computer security terminology.
2. Become familiar with basic cryptography concepts.
3. Learn to recognize potential phishing scams via email quickly.
4. Learn common attacks on servers and how to deal with them.

We developed four minigames to address each of these learning objectives individually:

1. Trivia covers a wide range of computer security topics and gradually increases in difficulty, which makes it an excellent tool for beginners and advanced users.
2. Key Hunter teaches players the basics of cryptography by introducing them to simple ciphers, an essential concept in computer security.
3. Phishing Frenzy trains the player to quickly recognize phishing attempts on an imaginary email account, preparing them for the inevitable slew of malicious emails they will experience in the real world.
4. Data Defenders simulates real-life attacks on servers, allowing the user to diagnose and respond to attacks, thus gaining experience with them.

The four included minigames target a wide range of skill levels and play styles, making the application an engaging and effective tool for learning computer security. Users will satisfy each learning objective by simply playing its corresponding game. How extensively they play is up to the discretion of the user. The longer the player engages with the application, the more they will learn to protect themselves. Every second of gameplay will enrich the player with the experience that will inevitably serve to protect them in the real world.

We have organized the remainder of the paper in the following manner: 2. Literature Review, 3. Tools and Technologies, 4. Minigames, 5. Discussion, and 6. Conclusion.

## 2. LITERATURE REVIEW

A study conducted at Utah State University suggests that students who pass tests demonstrating their knowledge of computer security have a lower rate of security-related issues [13]. The paper recommends incorporating network and security basics into the curriculum of undergraduate students, regardless of their majors. An additament of this study as a basis for aiming to protect the public can be extracted: general knowledge of computer security helps protect those who use computers and the internet.

Statistics shared by the top leadership of the U.S. Central Intelligence Agency identified that approximately 1,000 skilled specialists who can deter malicious attacks on government resources exist; however, to meet the Department of Defense's needs, 20,000 to 30,000 specialists are needed [14]. Such demand shows a staggering need for specialized workers in a field that most universities do not teach or promote [15]. This lack of cybersecurity professionals is not limited to the government [16]. Additionally, automated systems such as authentication using machine learning are not enough to make up for this discrepancy [17], [18], [19]. Thus, considering the evident lack of cybersecurity professionals, it would be prudent for individuals to learn to protect themselves.

There is a crucial distinction between learning that is done enthusiastically versus unenthusiastically – enthusiastic learning is conducive to higher retention [20]. Creating a web-based system that teaches cybersecurity concepts may be considered sufficiently relevant;

however, ensuring such a model has high efficacy would make it much more powerful [21]. Enthusiastic practice would allow the users to learn more effectively and thus better protect themselves [22]. One article details how a gamified approach to teaching cybersecurity increases the enjoyment of the process and turns it into a more effective learning tool [23]. The study explored the results of a role-playing quiz application on the Android platform. Designed to increase end-user awareness of cybersecurity concepts, it incentivized users to learn computer security concepts through its gamified approach. Subjects initially reported that their knowledge of computer security was not strong. After the sessions concluded, they established that the game helped increase their knowledge regarding password security and that the experience was enjoyable.

The literature on the effectiveness of game-based learning supports these conclusions [24]. Research has found game-based learning more effective and engaging than traditional learning methods – it is often seen as monotonous and dull, leading to poor learning results and ineffective learning culture [25]. Observations of game-based learning include the ability to promote faster acquisition of knowledge [26] and easier facilitation of intricate concepts such as science and mathematics [27].

## 3. TOOLS AND TECHNOLOGIES

The application's development included various tools and technologies, including Visual Studio Code, GitHub, Apache web server, HTML, CSS, JavaScript, and SQL. We used Visual Studio Code as the integrated development environment (IDE) for coding and GitHub as a collaboration tool for sharing files. The Apache web server allowed PhpMyAdmin and came with a built-in SQL IDE. HTML and CSS comprised the website's elements. Every unique page within the application (the front page, each minigame, etc.) had its own CSS and JavaScript file. The JavaScript provided the functional scaffolding for programming the minigames. JSON files stored a massive amount of data, with over 7000 lines of code dedicated to the Trivia game alone. All code features a neat style – consistent spacing, indentations, variable naming conventions, and logical and relevant organization styles. We constrained all HTML to one file to allow the one-page system described in section 3.1.

### 3.1. User Interface

Upon login, the application greets the user with a front page displaying their current high scores and ranks (where applicable) for each minigame (Figure 1). Every minigame is accessible from the front page with large accompanying logos, allowing effortless access for players of all ages and experience levels. Clear color themes for each game emerge once they become active. All pages are responsive and tactile, with easily clickable buttons and clear visuals. The website has animations smoothly incorporated throughout its contents, making for a visually pleasant experience for the user. All text's global minimum size is 16pt or 1em, allowing a viewing distance of 2 meters or about 6.5 feet.

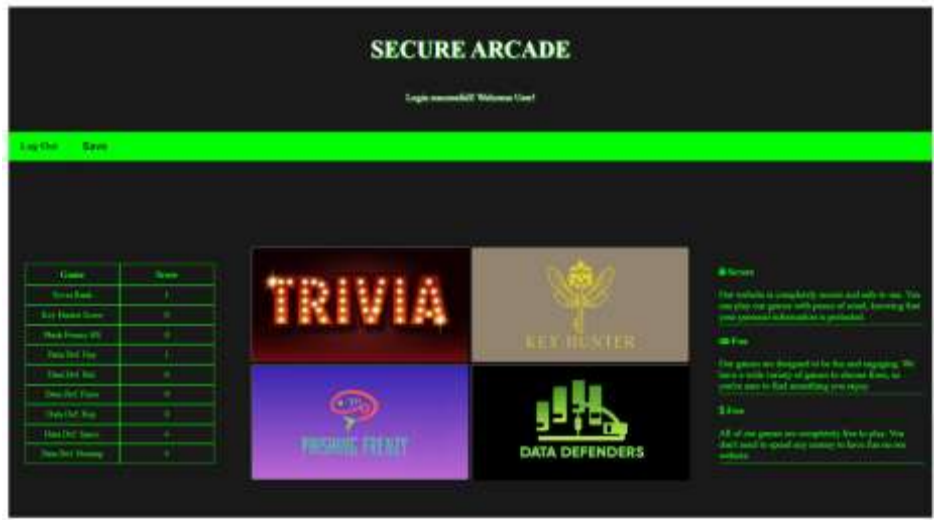

Figure 1. Front page of Secure Arcade, with a scores table (left), minigames (center), and basic info about the site (right)

## 3.2. Database Design

We created the database in phpMyAdmin using SQL code, based on effective paradigms established by previous work [28], [29], [30]. The database has five tables – admin, login, recover, stats, and user. Each table has a specific purpose and holds several types of data. Sensitive information is hashed and stored in the database to ensure the user's data is secure. Administrators cannot view a user's password or any other security data.

The "admin" table holds information about the administrators who have access to the database. This table contains data, such as the administrator's ID, username, password, and email address. This information grants access to the database.

The "login" table stores data related to user logins. It contains the user's ID, username, password, and email. This table verifies user login credentials and ensures secure access to the game platform. All the information stored in this table is encrypted to protect users from attacks.

The "recover" table assists users in recovering lost passwords. It contains the user's ID, username, password, and recovery email. It ensures that the user's data is kept secure and that only authorized users can recover their account information. All the information stored in this table is also encrypted to protect users from potential attackers.

The "stats" table is the most frequently used in the database. It stores data related to the minigames offered on the platform, including high scores, rankings, and other game-related information. The stats table retrieves information whenever a user completes playing a minigame and updates accordingly if there is a change in the retrieved values. As the platform evolves and we append more minigames, we will update this table to accommodate new data requirements.

Finally, the "user" table contains data about the platform's users. It includes the user's ID, nickname, email, username, and role. The role sends players to the game screen and administrators to the admin screen. Data in this table serves to pass and update values in either the stats table or the upcoming settings table.

Data is collected when players play each game, and once the player exits the minigame, the data gets permanently saved. The application does not preserve a session's data if a player leaves in the middle of a game.

# 4. MINIGAMES

We will now describe the four minigames featured in an online application designed to educate users on computer security concepts and improve their computer security skills. Each minigame caters to users of different skill levels and play styles.

## 4.1. Game 1: Trivia

The Trivia minigame is an effective tool for teaching computer security concepts to users of all levels. It provides a fun and engaging way for users to test their knowledge and improve their understanding of computer security topics. The minigame is divided into practice and ranked modes (Figure 2) to cater to the user's needs.

### 4.1.1. Practice Mode

In practice mode, users can select the number of questions they want to answer and choose to practice by topic or by rank (Figure 3). This feature is helpful for those who want to focus on specific areas of computer security that they may be struggling with. By practicing with targeted questions, they will improve their understanding and build confidence before moving on to more challenging content.

### 4.1.2. Ranked Mode

Ranked trivia games are always 25 questions long. Users start at rank 1. If users do well in the trivia game, they will rank up to a maximum potential level of 10. The questions become progressively more difficult with each increase in rank and delve deeper into computer security concepts – this way, users can gradually increase their knowledge and skill level without becoming overwhelmed or discouraged. Since there is no need for a selection process, once the user selects a rank, it asks if the user is ready to start (Figure 4).

A JSON database stores all the questions for Trivia. This database will be regularly updated to ensure that the questions remain relevant and up to date. We will conduct regular reviews for each question and their corresponding answers to ensure accuracy and relevance.

Trivia includes a diverse range of responses for each question. In addition to the traditional multiple-choice format (Figure 5), the game also incorporates true or false and multiple-correct-answer questions. The scoring mechanism follows a methodology reminiscent of the "Kahoot" platform, where the prompt's unveiling couples with a diminishing points allocation, incentivizing participants to respond promptly and accurately.

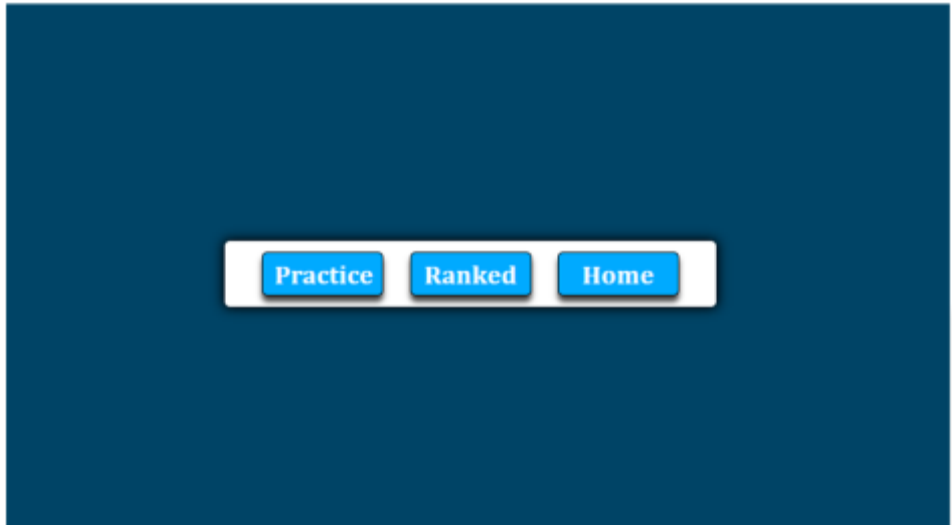

Figure 2. The front page of Trivia

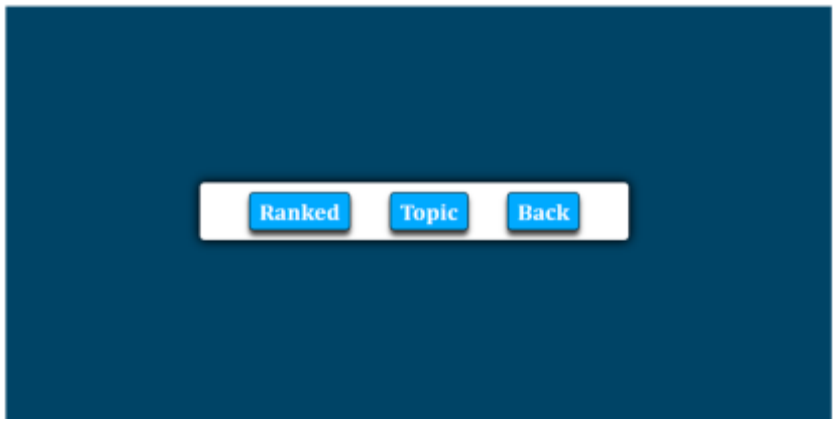

Figure 3. Players choose between a simulated ranked game or a specific topic in practice mode

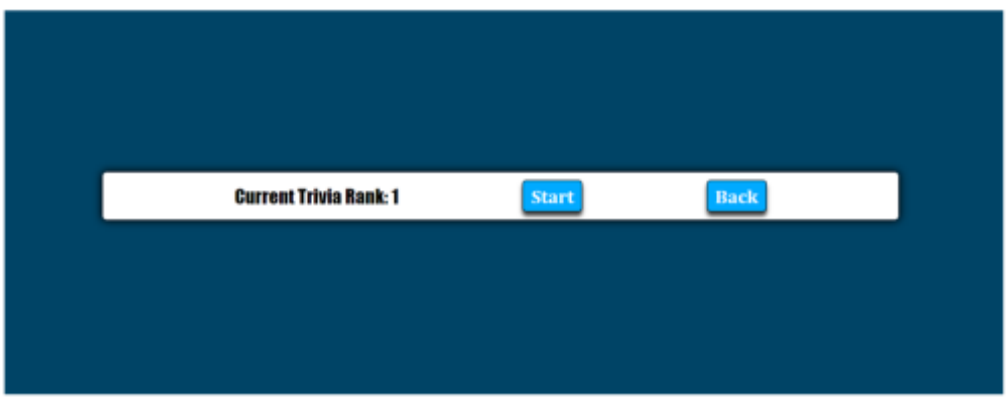

Figure 4. Ranked games provide fewer options and have the user begin with Rank 1 questions

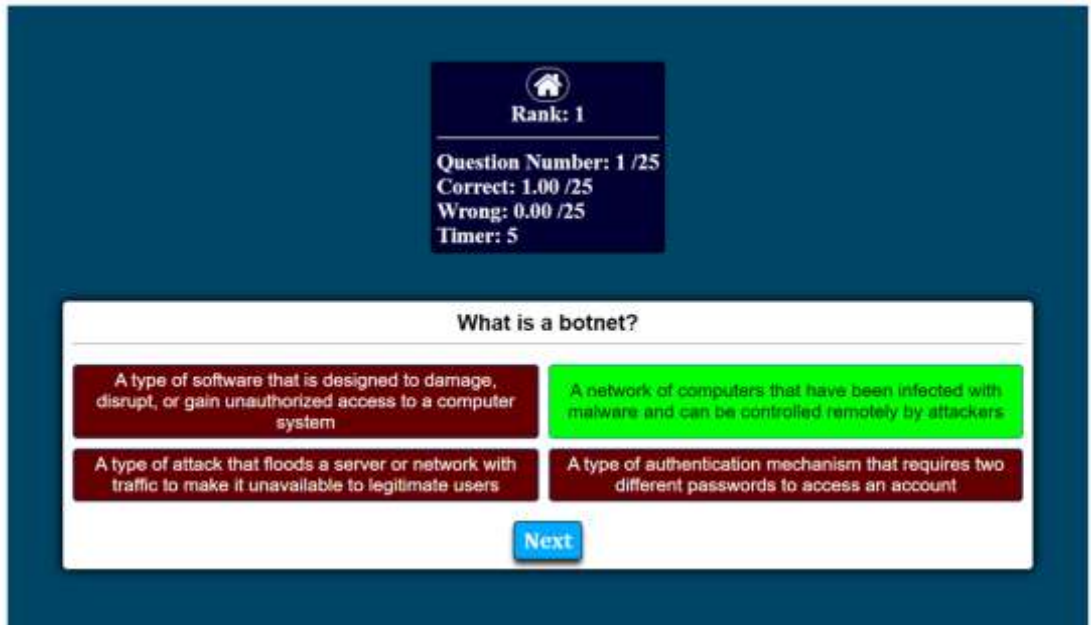

Figure 5. Gameplay of a ranked session

## 4.2. Key Hunter

Key Hunter is a crucial component of our online computer security application as it educates users on cryptography basics, an essential aspect of computer security. Cryptography is the practice of secure communication in the presence of third parties, also known as adversaries. Thus, cryptography allows us to encode information in a way that can only be read by those having the key to decipher it.

Key Hunter allows players to decipher various ciphers through a grid of buttons. The game has distinct difficulty levels (easy, medium, and hard) to cater to users of different skill levels (Figure 6). The "easy" mode introduces the players to three types of ciphers – the Pigpen cipher, Caesar cipher, and transposition cipher – and provides a basic understanding of how they work. The "medium" mode includes three more ciphers – a bash cipher, a zigzag cipher, and a Polybius cipher – which are more complex than the ones in the "easy" mode. "Hard" challenges users with three random ciphers selected from the previous six ciphers provided in "easy" and "medium."

Key Hunter offers a notable advantage by allowing the incorporation of new ciphers. The game uses a modular design where each cipher is independent of the others, allowing for easy addition or removal of ciphers. Additionally, the gameplay mechanics of Key Hunter are straightforward and intuitive. The game presents players with a grid of buttons and an encrypted message (Figure 7). The objective is to decrypt the message, thereby revealing the precise coordinates of the correct button. However, if the player mistakenly selects an incorrect coordinate, it promptly turns red, eliminating one of their five "attempts."

A time limit of 5 minutes adds an element of constraint and challenge, encouraging quick thinking and strategic decision-making. Moreover, the game interface features five tabs thoughtfully positioned on the side, in a top-to-bottom order: the dictionary tab, message tab, notes tab, question tab, and home button (Figure 8). The dictionary tab provides a "hint" system for the various potential encryption options, showing how each cipher works. The message tab (Figure 9) offers descriptions of each cipher to aid players should they require additional explanation. The notes tab affords the freedom to record personalized information or insights in a dedicated text box. Lastly, the question tab provides a descriptive overview of the game, ensuring players remain engaged and well-informed throughout their Key Hunter experience. Ultimately, the objective of Key Hunter is to achieve a high score that is then meticulously recorded within the SQL database, adding a competitive and rewarding dimension to the gameplay.

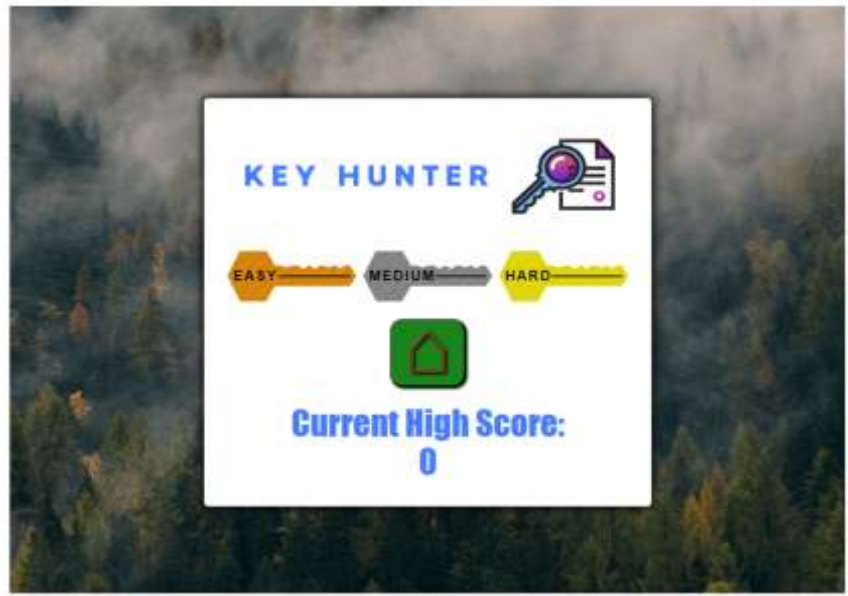

Figure 6. The login screen of Key Hunter allows a player to choose between three difficulties

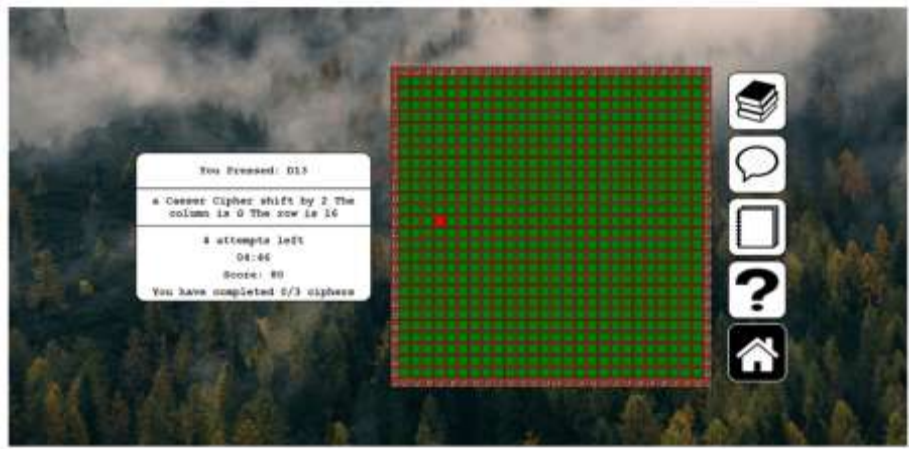

Figure 7. Basic gameplay of Key Hunter. The Caesar Cipher is the first random cipher chosen in this instance (with the red coordinate marking an incorrect guess)

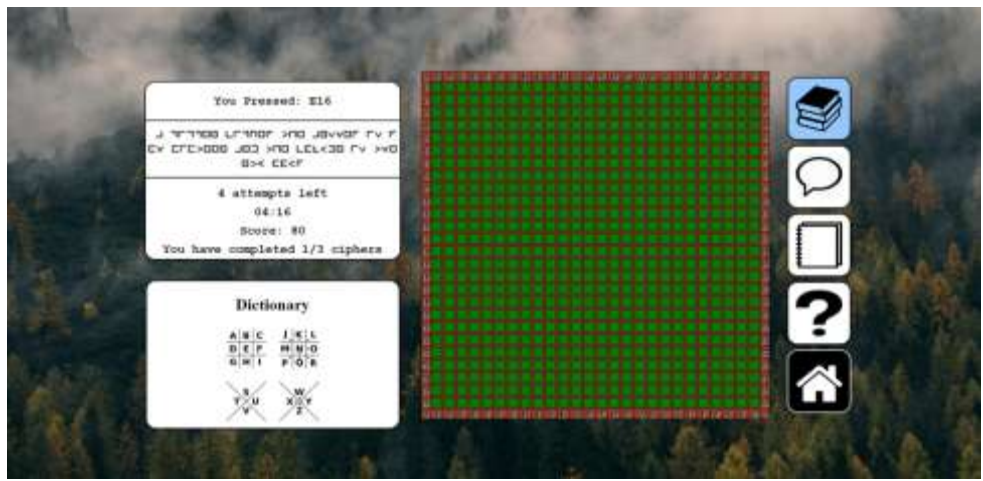

Figure 8. Gameplay with the dictionary tab open.

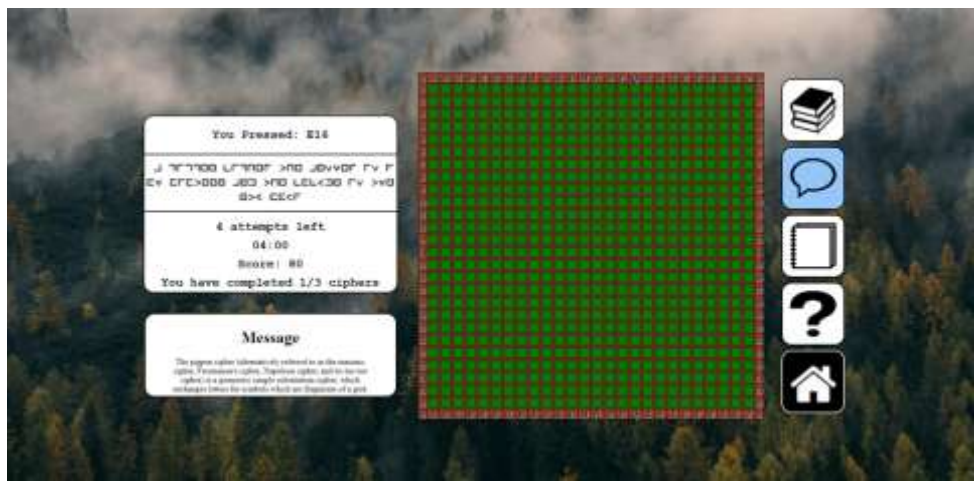

Figure 9. Gameplay with the messages tab open

### 4.3. Phishing Frenzy

Phishing is one of the most common cyber-attacks that can lead to serious security breaches. The attackers usually send emails that appear to be from a legitimate source to trick the recipient into clicking on a link, opening an attachment, or entering personal information. These emails can look very convincing, often mimicking the branding and language of a trusted company or individual. The consequences of failing a phishing attack can range from identity theft and monetary loss to unauthorized access to sensitive information.

In today's digital age, everyone is a potential target of phishing attacks, regardless of age, occupation, or level of computer literacy. Therefore, it is essential to educate people about how to recognize and avoid phishing attempts. Simulating real-life scenarios helps players develop their critical thinking and analytical skills and learn how to identify and respond to phishing emails – and Phishing Frenzy is a game designed to do just that.

Phishing Frenzy utilizes a JSON database filled with potential emails to generate hundreds of possibilities. While playing the game, players will encounter numerous phishing techniques and tricks, making for a realistic and challenging experience. The game also provides feedback on each email, explaining why it is a phishing attempt or a legitimate message, and offers tips on how to spot the signs of phishing. This system allows smooth implementation of admin changes, where the administrator can add or remove messages.

After the player selects a difficulty level in Phishing Frenzy (Figure 10), players become immersed in an interface resembling an email application (Figure 11). The screen features a distinct sender address and message randomly selected from the JSON file. The player's primary objective is to discern whether the presented email is a "Legitimate" message or a "Phishing" attempt. Upon making their determination, by pressing the corresponding button, the email is promptly sorted into its inbox – legitimate messages on the left and phishing messages on the right. Users can hover their cursor over the inbox tab to facilitate a comprehensive understanding of each message's legitimacy or phishing nature. This action unveils the email's content and explains why it was legitimate or a phishing attempt (Figure 12).

Accurate identification of messages is rewarded with points, accompanied by the display of a green checkmark or green fish symbol within the corresponding inbox. However, incorrectly identifying an email means eliminating one of the player's three lives and is accompanied by a red checkmark or a "red fish" symbol within the corresponding inbox.

In Phishing Frenzy, the goal is to accumulate as many points as possible within one minute. The high scores achieved by players are recorded and stored within the game's database, fostering a sense of competition and enabling players to strive for excellence in their subsequent attempts.

Phishing Frenzy is a valuable tool for educators and organizations who want to promote awareness about computer security and protect their employees or students from phishing attacks. By making the learning experience engaging and interactive, the game can raise awareness and assist in fostering a culture of security. With the increasing prevalence of cyber-attacks, it is more important than ever to equip people with the skills and knowledge they need to stay safe online.

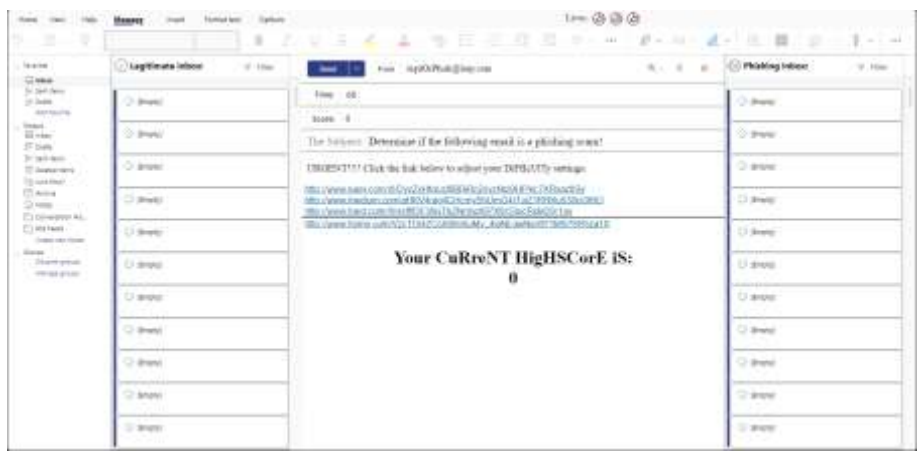

Figure 10. Phishing Frenzy home screen

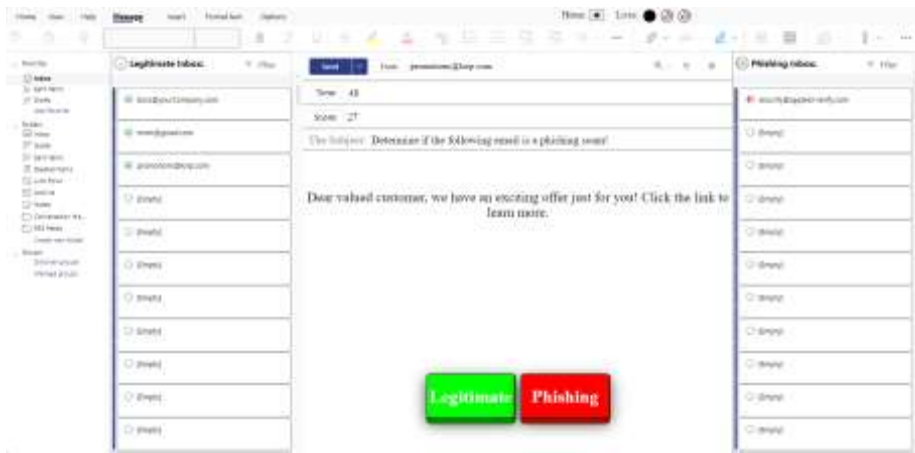

Figure 11. Basic gameplay of Phishing Frenzy (One incorrect answer marked on the right, three correct answers on the left, "lives" on Top)

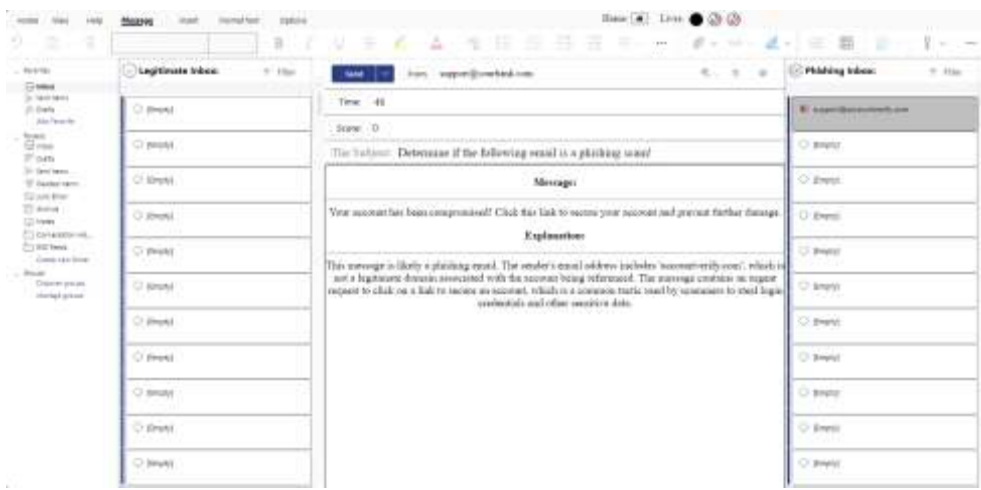

Figure 12. Highlighting a wrong answer explains why a message may or may not be phishing

## 4.4. Data Defenders

Data Defenders offers a unique approach to teaching players about computer security. In the minigame, the player oversees four servers hosting four websites each. It involves real-time management of the servers throughout a simulated day cycle. This approach offers an immersive and engaging way for players to learn about diverse types of cyberattacks and how to respond to them. The game provides the user with various data sources, including server performance metrics, security cameras, and messages from website owners, providing the player with a realistic simulation of what it is like to be responsible for maintaining and securing a network of websites. Players can view these various data sources and clues to help diagnose and respond to attacks on the servers using the minigame's multi-tab interface. The game also incorporates a reputation score, determining how successful the player is at protecting their network. This aspect of the game helps to reinforce the importance of computer security in a real-world context.

Upon opening the minigame, the application greets the player with a screen resembling a home computer's desktop (Figure 13). The player can begin a day by selecting the "Start New Day" button. Throughout a "day," websites or servers will have a simulated attack in one of six forms: a DoS (Denial of Service) Attack, Malware Attack, DNS attack, Insider Intrusion, SQL Injection, or a USB Drop attack. The player must then look through the available tabs at the bottom of the screen to get clues pointing toward what type of attack is happening. The "Websites" tab (Figure 14) displays a table of websites showing all folders with files and their pathing information. The "Servers" tab (Figure 15) shows the health and performance of the four sites inside each server – the player can view system calls and the IP addresses of connections. The "Sec. Cams" tab (Figure 16) features security cameras showing if there is an insider intruder or USB drop attack. The "Messages" tab (Figure 17) is a messages tab that updates with messages from the owners of the websites that tell you they are experiencing issues.

Once the player has diagnosed the attack, they will report it using all pertinent information and must answer four questions about the correct course of action using the "Report" tab (Figure 18). How well the player does on these multiple-choice questions affects their "reputation." Reputation determines if more websites join a player's hosting services. The more websites you have, the more money you get, which allows the player to upgrade their servers. To receive the most reputation points, players must diagnose the type of attack and respond appropriately, including answering the questions correctly and speedily. This aspect offers a competitive element that keeps players engaged and motivated to learn more.

Data Defenders is an excellent tool for teaching players about computer security. By providing a realistic simulation of several types of attacks and offering various data avenues to diagnose and respond, the game becomes an engaging and effective method for players to learn about computer security. Furthermore, real-world scenarios and a competitive reputation system help reinforce the importance of computer security in a practical, meaningful way.

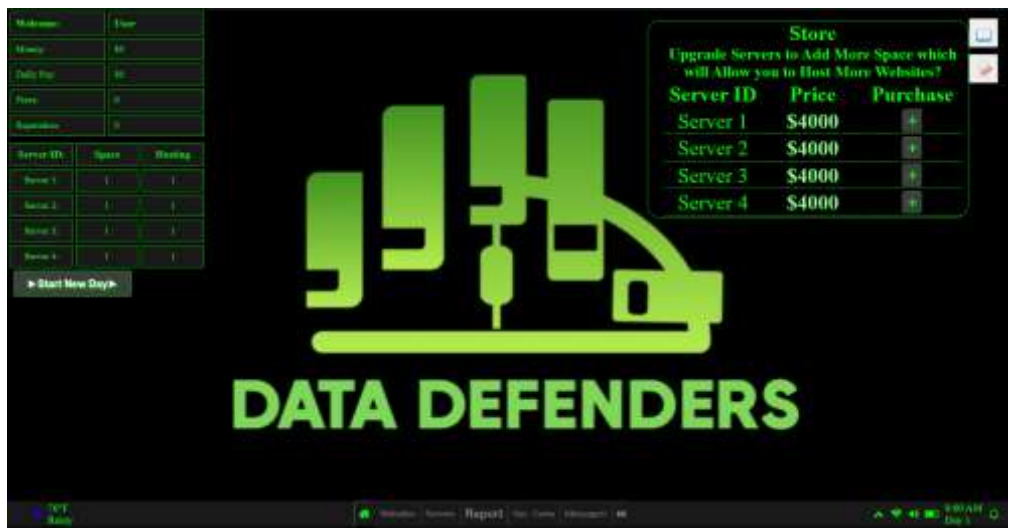

Figure 13. The main page of Data Defenders

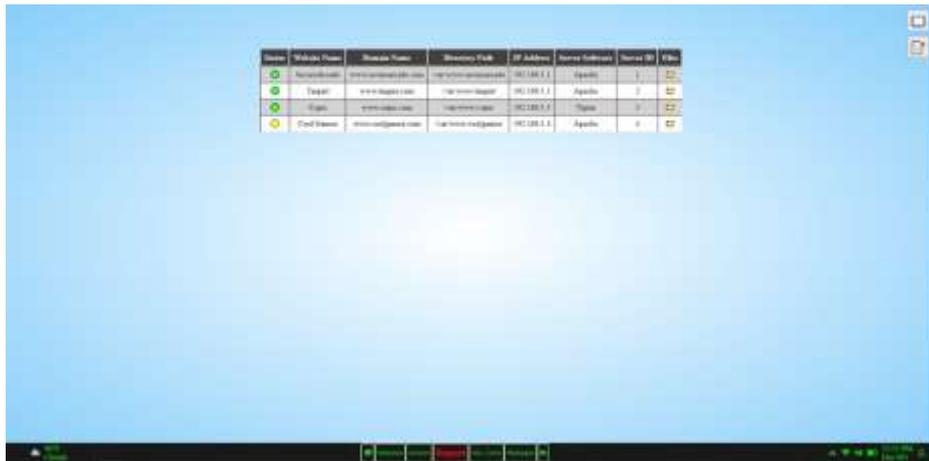

Figure 14. The "Websites" tab with one of the domains experiencing an attack

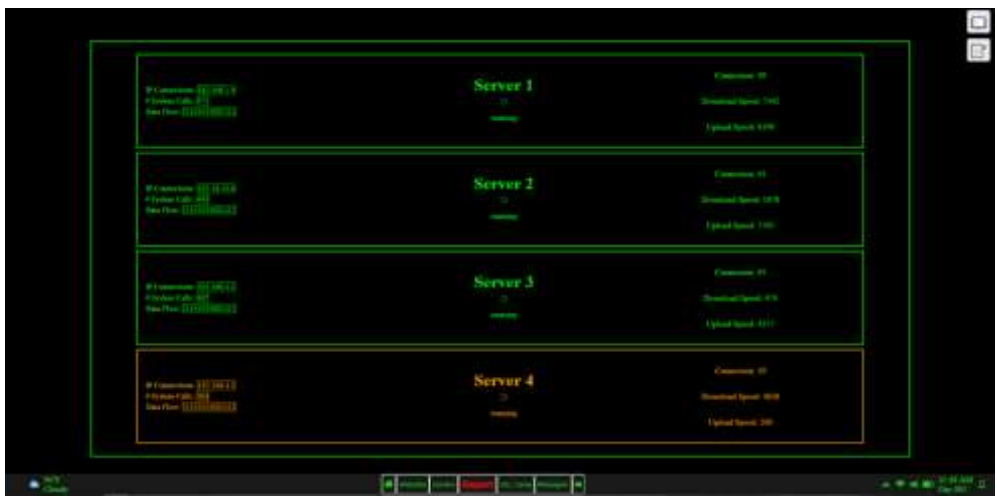

Figure 15. The "Servers" tab with server 4 experiencing a downgrade in performance due to a prolonged attack

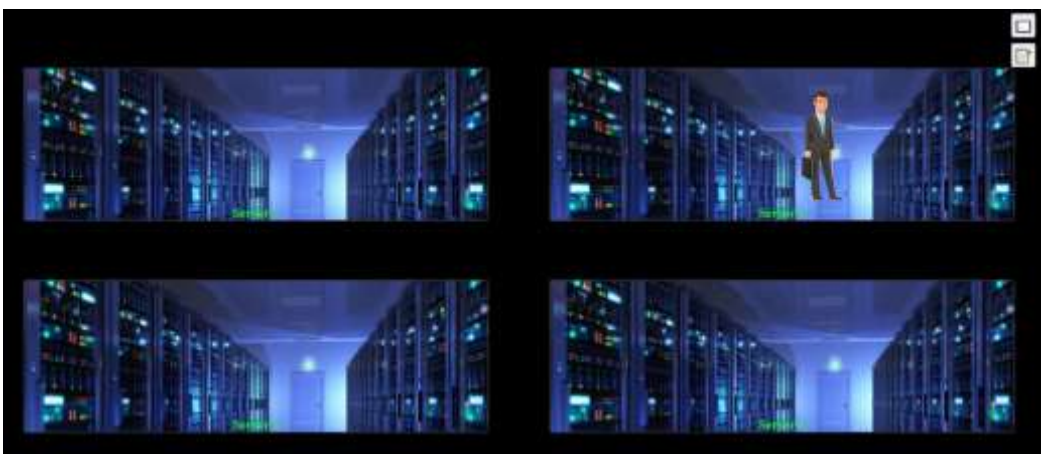

Figure 16. The "Sec. Cams" tab with an insider intruder attack occurring in server room 2

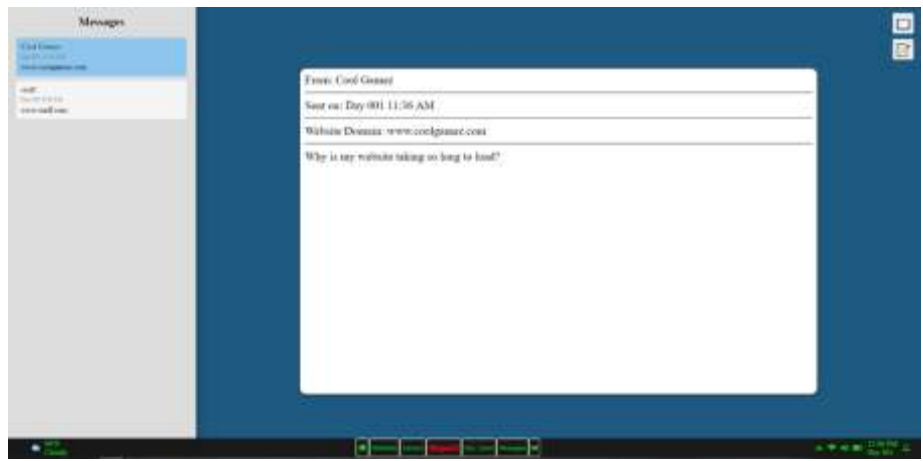

Figure 17. The "Messages" tab featuring a message from a domain alluding to an attack

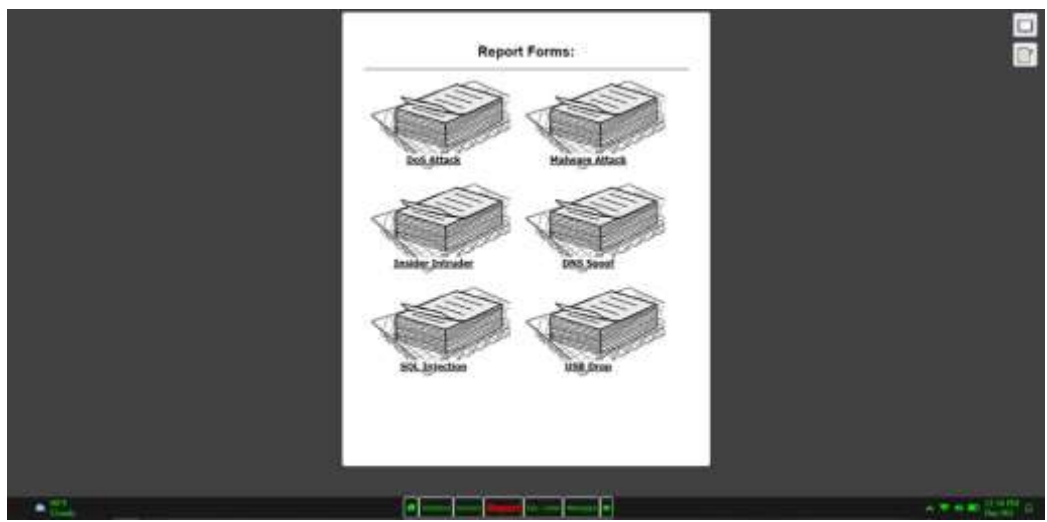

Figure 18. The "Report" tab

## 5. DISCUSSION

In this project, we expressed the effectiveness of a website that utilizes minigames to teach various aspects of computer security. The website aims to provide an engaging and interactive learning experience, equipping them with essential knowledge and skills to navigate the complex landscape of computer security threats. Through the application's breakdown of computer security concepts into unique game scenarios, users can actively apply their knowledge and practice real-world computer security strategies.

The importance of such a website lies in its ability to address the growing need for computer security education [31]. As technology continues to evolve and permeate every aspect of our lives, the threat landscape expands, making individuals and organizations vulnerable to computer security attacks. By incorporating gamification elements, the website offers an accessible and enjoyable platform for users to learn about different facets of computer security, such as secure coding, network defense, data privacy, and social engineering.

However, we must acknowledge certain limitations in this approach. Firstly, the application's efficacy in teaching computer security concepts will vary depending on the design and quality of the minigames. We must carefully ensure that the games accurately represent real-world scenarios and provide realistic challenges. Additionally, we must consider the website's relevance to the rapidly evolving computer security landscape as new threats and attack techniques constantly emerge [32]. Thus, continuous updates and enhancements to the mini-games and content are paramount. Regular reviews and collaborations with cybersecurity experts would help ensure the website remains up to date with the latest trends and best practices. Furthermore, incorporating adaptive learning techniques, such as personalized feedback and tailored content, would help users of different skill levels and backgrounds progress at their own pace. Finally, new updates and their requirement specifications should be scrutinized to mitigate any potential weaknesses during development before an update is publicly released [33], [34].

## 6. CONCLUSION

Developing an online application that utilizes a fun and interactive minigame system to teach computer security concepts is an excellent step toward increasing online vigilance, which helps individuals protect themselves and their digital assets. With the increasing number of devices connected to the internet and the corresponding rise in cyber threats, computer security has become crucial in safeguarding privacy and supporting a tranquil modern life.

The four minigames included in Secure Arcade target different skill levels and play styles, making it accessible to users of all levels. These minigames provide users opportunities to test their knowledge and improve their understanding of computer security topics, such as cryptography and recognizing phishing attempts, which are essential concepts in computer security.

Overall, this online application delivers a unique and innovative method of teaching computer security concepts to people of all skill and knowledge levels. The application's fun and interactive minigame system makes it an excellent tool for promoting higher online vigilance and helping individuals protect themselves and their digital assets.